\documentclass[aps,prb,showpacs,showkeys]{revtex4}
\usepackage{amsfonts}
\usepackage{amssymb}
\usepackage{amsmath}
\usepackage{graphicx}
\usepackage{amsbsy}
\usepackage{color}

\newcommand{\bq}{\begin{eqnarray}}
\newcommand{\eq}{\end{eqnarray}}
\newcommand{\bqn}{\begin{eqnarray*}}
\newcommand{\eqn}{\end{eqnarray*}}

\begin{document}

\preprint{}
\title{Local orientational ordering in fluids of spherical molecules \\
with dipolar-like anisotropic adhesion. }
\author{Domenico Gazzillo, Riccardo Fantoni and Achille Giacometti}
\affiliation{Dipartimento di Chimica Fisica, Universit\`a di Venezia, S. Marta DD 2137,
I-30123 Venezia, Italy}

\begin{abstract}
We discuss some interesting physical features stemming from our previous
analytical study of a simple model of a fluid with dipolar-like interactions
of very short range in addition to the usual isotropic Baxter potential for
adhesive spheres. While the isotropic part is found to rule the global
structural and thermodynamical equilibrium properties of the fluid, the
weaker anisotropic part gives rise to an interesting short-range local ordering 
of nearly spherical condensation clusters, containing short portions of chains
having nose-to-tail parallel alignment which runs antiparallel to adjacent similar chains.
\end{abstract}

\date{\today}
%\startpage{1}
\maketitle

%%%%%%%%%%%%%%%%%%%%%%%%%%%%%%%%%%%%%%%%%%%%%%%%%%%%%%%%%%%%%%%%%%%%%%%%%%%%%%
%\section{Introduction}
%%%%%%%%%%%%%%%%%%%%%%%%%%%%%%%%%%%%%%%%%%%%%%%%%%%%%%%%%%%%%%%%%%%%%%%%%%%%%%
%\label{sec:1}

Even simple hard sphere fluids display a non-trivial phase diagram, as a
function of the packing fraction, which can be experimentally probed and
theoretically interpreted\cite{Zaccarelli09}. Softening the potential and/or
increasing its range, leads to a remarkably richer phase diagram which has
attracted considerable attention recently (see e.g. Ref. \onlinecite{Zaccarelli08} for
a recent review). Yethiraj and van Blaaderen \cite{Yethiraj03} have
discussed how it is experimentally possible to tune the interactions from
hard sphere to soft and dipolar ones. More recently, Lu \textit{et al.} \cite%
{Lu08} have shown that, contrary to an intuitive expectation, gelation of
particles with short-range attractions is intimately connected with its
equilibrium phase diagram.

It is widely believed that the addition of a long-range repulsion to a
short-range attraction inhibits phase separation, by promoting the formation
of an equilibrium gel. The same mechanism can be achieved by reducing the
probability of forming a bulk liquid using the concept of limited-valency
and/or patchy particles \cite{Zaccarelli08}. This idea has been recently
explored by a number of authors, who have used the so-called Kern and
Frenkel model with circular adhesive patches (of non-vanishing
area), or that with short-ranged attractive point-sites on the
surface of hard spheres \cite%
{Kern03,Fantoni07,Liu07,Foffi07,Bianchi08,Gogelein08,Tavares09}.

In spite of their usefulness, the above models share a common shortcoming on
the discontinuous dependence of the potential on the particle orientations,
which makes them very difficult to investigate from a theoretical point of
view. This drawback is not present in molecular interactions where this
dependence is continuous, as for instance in dipolar interactions \cite%
{Wertheim71}, a case which is particularly interesting for various reasons.
First, because of their widespread appearance in colloidal suspensions, such
as ferrofluids, which have important practical applications. In addition,
recent studies \cite{Tlusty00,Camp00,Ganzenmuller07} have shown the
existence of a significant influence, in the equilibrium properties of the
fluid, of chain-like aggregation characteristic of the dipolar interaction,
which strongly competes with a stable fluid-fluid phase separation.

Motivated by this features, in this paper we then take an extreme
alternative of considering a tail with dipolar-like anisotropy combined with
a very short-range attraction. The latter is patterned after the well-known
Baxter's sticky hard sphere (SHS) potential, where attraction occurs only at
contact \cite{Baxter68}. Building upon our previous, almost fully
analytical, study on this model within the Percus-Yevick closure
with orientational linearization (PY-OL) \cite{Gazzillo08}, we discuss here
some additional interesting features on the local ordering properties, which
were not accounted for in our previous work.

In the same spirit of Baxter's isotropic counterpart \cite{Baxter68}, the
model is defined by the following Mayer function \cite{Gazzillo08} 
\begin{equation}
f(1,2)=f_{\mathrm{HS}}(r)+t\ \epsilon (1,2)\ \sigma \delta \left( r-\sigma
\right) ,  \label{model:eq1}
\end{equation}%
where $f_{\mathrm{HS}}(r)=\Theta \left( r-\sigma \right) -1$ is its hard
sphere (HS) contribution, $\Theta $ the Heaviside step function ( $\Theta
(x<0)=0$, $\Theta (x>0)=1$ ), and the Dirac delta function $\delta \left(
r-\sigma \right) $ ensures that the adhesive interaction occurs only at
contact ($\sigma $ is the HS diameter). The symbol $i\equiv \left( \mathbf{r}%
_{i},\Omega _{i}\right) $ (with $i=1,2$) denotes both the position $\mathbf{r%
}_{i}$ of the molecular center and the orientation $\Omega _{i}$ which
combines the usual polar and azimuthal angles $\left( \theta _{i},\varphi
_{i}\right) $. Thus we have: $(1,2)=(\mathbf{r}_{12},\Omega _{1},\Omega
_{2})=(r,\widehat{\mathbf{r}}_{12},\Omega _{1},\Omega _{2})=(r,\Omega
_{r},\Omega _{1},\Omega _{2})$, with $\mathbf{r}_{12}=\mathbf{r}_{2}-\mathbf{%
r}_{1}$, $r=|\mathbf{r}_{12}|$, and $\Omega _{r}$ being the solid angle
associated with $\widehat{\mathbf{r}}_{12}=\mathbf{r}_{12}/r$. Moreover, $t$
is the stickiness parameter, equal to $\left( 12\tau \right) ^{-1}$ in
Baxter's original notation \cite{Baxter68}, which measures the strength of
surface adhesion and increases with decreasing temperature.

Finally, the angular dependence of the surface adhesion is expressed through
the angular factor%
\begin{equation}
\epsilon (1,2)=1+\alpha D(1,2),  \label{model:eq2}
\end{equation}%
including the dipolar function%
\begin{equation}
D(1,2)=D(\Omega _{r},\Omega _{1},\Omega _{2})=3(\hat{\mathbf{r}}\cdot 
\mathbf{u}_{1})(\hat{\mathbf{r}}\cdot \mathbf{u}_{2})-\mathbf{u}_{1}\cdot 
\mathbf{u}_{2}  \label{model:eq3}
\end{equation}%
Here and in the following, the unit vector $\mathbf{u}_{i\text{ }}$
represents the orientation $\Omega _{i}$ of molecule $i$, while $\hat{%
\mathbf{r}}$ coincides with $\hat{\mathbf{r}}_{12}=-\hat{\mathbf{r}}_{21}$.

The anisotropic function $\epsilon (1,2)$, which has the same symmetry as
the dipolar interaction, modulates the sticky attraction. The requirement $%
\epsilon (1,2)\geq 0$ along with $-2\leq D(1,2)\leq 2$ enforce the limits $%
0\leq \alpha \leq \frac{1}{2}$ on the anisotropy degree. This range
corresponds to the surface interaction always being attractive. In the
isotropic case, one has $\alpha =0$ and $\epsilon (1,2)=1$.

As convolutions of Mayer functions generate correlation functions with a
more complex angular dependence \cite{Wertheim71}, it is necessary to
consider also the angular function 
\begin{equation}  \label{model:eq4}
\Delta (1,2)=\mathbf{u}_{1}\cdot \mathbf{u}_{2}\text{ },
\end{equation}%
whose limits of variation are clearly $-1\leq \Delta (1,2)\leq 1$.

We note the difference between the dipolar anisotropic adhesion introduced
here and the anisotropy belonging to the class of uniform circular 'sticky
patches' \cite{Jackson88,Ghonasgi95,Sear99,Mileva00,Kern03,Zhang05,Fantoni07}%
. In the latter case, the strength of adhesion is uniform, independent of
the contact point inside an attractive patch, whereas in our model the value
of \ the anisotropic correction $\alpha t\ D(1,2)$ changes with the position
of the contact point. Moreover, $D(1,2)$ can assume both positive and
negative values, depending on the molecular orientations. Consequently, the
strength of adhesion between two particles $1$ and $2$ at contact depends --
in a continuous way -- on the relative orientation of $\mathbf{u}_{1}$ and $%
\mathbf{u}_{2}$ as well as on the unit vector $\widehat{\mathbf{r}}_{12}$ of
the intermolecular distance. The orientations with $D(1,2)>0$, and thus with 
$\epsilon (1,2)=1+\alpha \ D(1,2)>1$, correspond to an attraction stronger
than the isotropic one (given by $\epsilon (1,2)=1$), whereas the
configurations with $D(1,2)<0$, and thus with $\epsilon (1,2)<1$, are
characterized by a weaker attraction, which can even reduce to zero (HS
limit) in the case of highest anisotropy admissible in the present model,
i.e. $\alpha =1/2$.

In particular, we shall focus on a set of \textit{parallel and antiparallel}
configurations with $\Delta (1,2)=\mathbf{u}_{1}\cdot \mathbf{u}_{2}=1$, and 
$\Delta (1,2)=-1$ respectively. The surface adhesion reaches its maximum
value when $\mathbf{u}_{1}=\mathbf{u}_{2}=$ $\widehat{\mathbf{r}}_{12}$,
which yields $D(1,2)=2$ and $\epsilon (1,2)=1+2\alpha $ (head-to-tail
parallel configuration). On the contrary, the stickiness is minimum, and
vanishes for $\alpha =1/2$, when $\mathbf{u}_{1\text{ }}=-$ $\mathbf{u}_{2}=%
\widehat{\mathbf{r}}_{12}$, which corresponds to $D(1,2)=-2$ and $\epsilon
(1,2)=1-2\alpha $ (head-to-head or tail-to-tail antiparallel
configurations). The intermediate case of \textit{orthogonal }configuration (%
$\mathbf{u}_{2}$ perpendicular to $\mathbf{u}_{1}$) corresponds to $%
D(1,2)=0, $ which is equivalent to the isotropic SHS interaction.

Introducing the orientational average $\langle \ldots \rangle _{\mathbf{u}%
}=\left( 4\pi \right) ^{-1}\int d\mathbf{u}\ldots $ we note the following
results 
\begin{equation}
\begin{array}{c}
\left\langle \Delta (1,2)\right\rangle _{\mathbf{u}_{1},\mathbf{u}%
_{2}}=0\qquad \left\langle D(1,2)\right\rangle _{\mathbf{u}_{1},\mathbf{u}%
_{2}}=0 \\ 
\\ 
\left\langle \Delta (1,2)D(1,2)\right\rangle _{\mathbf{u}_{1},\mathbf{u}%
_{2}}=0\text{ \ \ \ \ and \ \ \ \ \ }\left\langle D^{2}(1,2)\right\rangle _{%
\mathbf{u}_{1},\mathbf{u}_{2}}=\frac{2}{3}%
\end{array}
\label{model:eq5}
\end{equation}

In a previous paper (hereafter referred to as Ref. I) \cite{Gazzillo08}, we
have analytically solved for this model the Percus-Yevick integral equation
with an orientational linearization.

We here recall the main results, referring to Ref. I for details. We start
with the molecular Ornstein-Zernike (OZ) integral equation for homogeneous
fluids

\begin{equation}
h(1,2)=c(1,2)+\rho \int d\mathbf{r}_{3}\ \left\langle \ c(1,3)\ h(3,2)\
\right\rangle _{\mathbf{u}_{3}}\   \label{summary:eq1}
\end{equation}%
where $h(1,2)$ and $c(1,2)$ are the total and direct correlation functions,
respectively, and $\rho $ is the number density.

Any angle-dependent correlation function $F(1,2)$ could be expanded in a
basis of rotational invariants \cite{Gray84}, whose first few terms are 
\begin{equation}
F(1,2)=F_{0}(r)+F_{\Delta }(r)\Delta (1,2)+F_{D}(r)D(1,2)+\cdots \text{\ ,}
\label{summary:eq2}
\end{equation}%
\noindent We stop at the linear terms, assuming \cite{Wertheim71,Gazzillo08}
that the angular basis $\left\{ 1,\Delta ,D\right\} $ is sufficient for our
purposes.

The \textit{PY-OL closure} \cite{Gazzillo08} is a combination of the PY
closure, i.e. $c^{\mathrm{PY}}=f\left( 1+\gamma \right) $, with the linear
expansion of $\gamma \equiv h-c$ given by $\gamma ^{\mathrm{OL}}(1,2)=\gamma
_{0}(r)+\gamma _{\Delta }(r)\Delta (1,2)+\gamma _{D}(r)D(1,2)$, which also
neglects the $D\Delta $ and $D^{2\text{ }}$terms stemming from the product $%
f\gamma $. This leads to 
\begin{equation}
c^{\mathrm{PY-OL}}(1,2)=c_{0}(r)+c_{\Delta }(r)\Delta (1,2)+c_{D}(r)D(1,2)
\end{equation}%
\begin{equation}
\left\{ 
\begin{array}{c}
c_{0}(r)=f_{\mathrm{HS}}(r)[1+\gamma _{0}(r)]+\Lambda _{0}\ \sigma \delta
\left( r-\sigma \right)  \\ 
\\ 
c_{\Delta }(r)=f_{\mathrm{HS}}(r)\gamma _{\Delta }(r)+\Lambda _{\Delta }\
\sigma \delta \left( r-\sigma \right)  \\ 
\\ 
c_{D}(r)=f_{\mathrm{HS}}(r)\gamma _{D}(r)+\Lambda _{D}\sigma \delta \left(
r-\sigma \right) 
\end{array}%
\right.   \label{summary:eq3}
\end{equation}%
\begin{equation}
\Lambda _{0}=t\ [1+\gamma _{0}(\sigma )],\text{ \ \ \ \ \ \ }\Lambda
_{\Delta }=t\ \gamma _{\Delta }(\sigma ),\text{ \ \ \ \ \ \ }\Lambda _{D}=t\
\gamma _{D}(\sigma )+\alpha \Lambda _{0}.
\end{equation}%
\noindent 

The solution of the OZ equation with the above closure then yields the
approximate pair distribution function 
\begin{equation}
g^{\mathrm{PY-OL}}(1,2)=1+h^{\mathrm{PY-OL}}(1,2)=g_{0}(r)+h_{\Delta
}(r)\Delta (1,2)+h_{D}(r)D(1,2)  \label{gPYOL}
\end{equation}%
\begin{equation}
\left\{ 
\begin{array}{c}
g_{0}(r)=e_{\mathrm{HS}}(r)[1+\gamma _{0}(r)]+\Lambda _{0}\ \sigma \delta
\left( r-\sigma \right)  \\ 
\\ 
h_{\Delta }(r)=e_{\mathrm{HS}}(r)\gamma _{\Delta }(r)+\Lambda _{\Delta }\
\sigma \delta \left( r-\sigma \right)  \\ 
\\ 
h_{D}(r)=e_{\mathrm{HS}}(r)\gamma _{D}(r)+\Lambda _{D}\sigma \delta \left(
r-\sigma \right) 
\end{array}%
\right.   \label{summary:eq4}
\end{equation}%
where $g_{0}(r)=1+h_{0}(r)$, and $e_{\mathrm{HS}}(r)=1+f_{\mathrm{HS}}(r)$
is the HS Boltzmann factor.

The first term in Eqs.~(\ref{summary:eq3}) corresponds to the well known
isotropic Baxter's sticky hard sphere solution \cite{Baxter68} and the OZ
equation and this closure constitute a self-contained system. The remaining
two have a similar form, but they depend in a non-trivial way upon the
isotropic term (see Ref. I for details).

It is instructive to consider the behavior of the $g(12)$ assuming that $%
\widehat{\mathbf{r}}_{12}\cdot \mathbf{u}_{1}=1$. We focus on a generic
reference particle 1, with fixed position $\mathbf{r}_{1}$ and orientation $%
\mathbf{u}_{1}$, and consider a particle 2 located along the straight
half-line which originates from $\mathbf{r}_{1}$ and has the same direction
as $\mathbf{u}_{1}$ (polar axis). Imagine that 2 has fixed distance $r$ from
1, but can assume all possible orientations $\mathbf{u}_{2}$, which -- by
axial symmetry -- can be described by the single angle $\theta _{12}=\cos
^{-1}(\mathbf{u}_{1}\cdot \mathbf{u}_{2})$. Consequently, $g(1,2)$ reduces
to: $g(r,\theta _{12})=g_{0}(r)+\left[ h_{\Delta }(r)+2h_{D}(r)\right] (%
\mathbf{u}_{1}\cdot \mathbf{u}_{2})$.

\begin{figure}[tbph]
\begin{center}
\includegraphics[width=12cm]{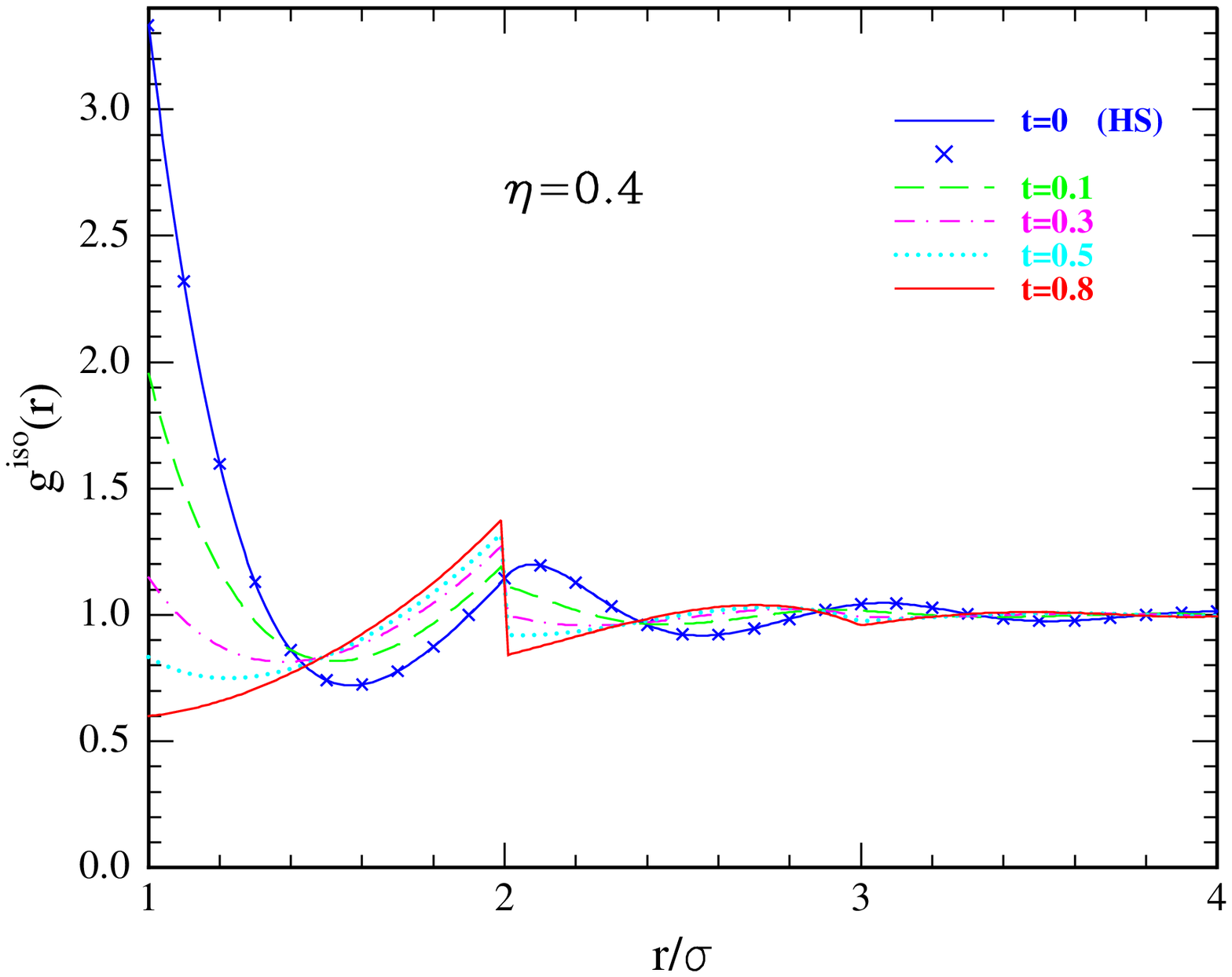} %
\includegraphics[width=12cm]{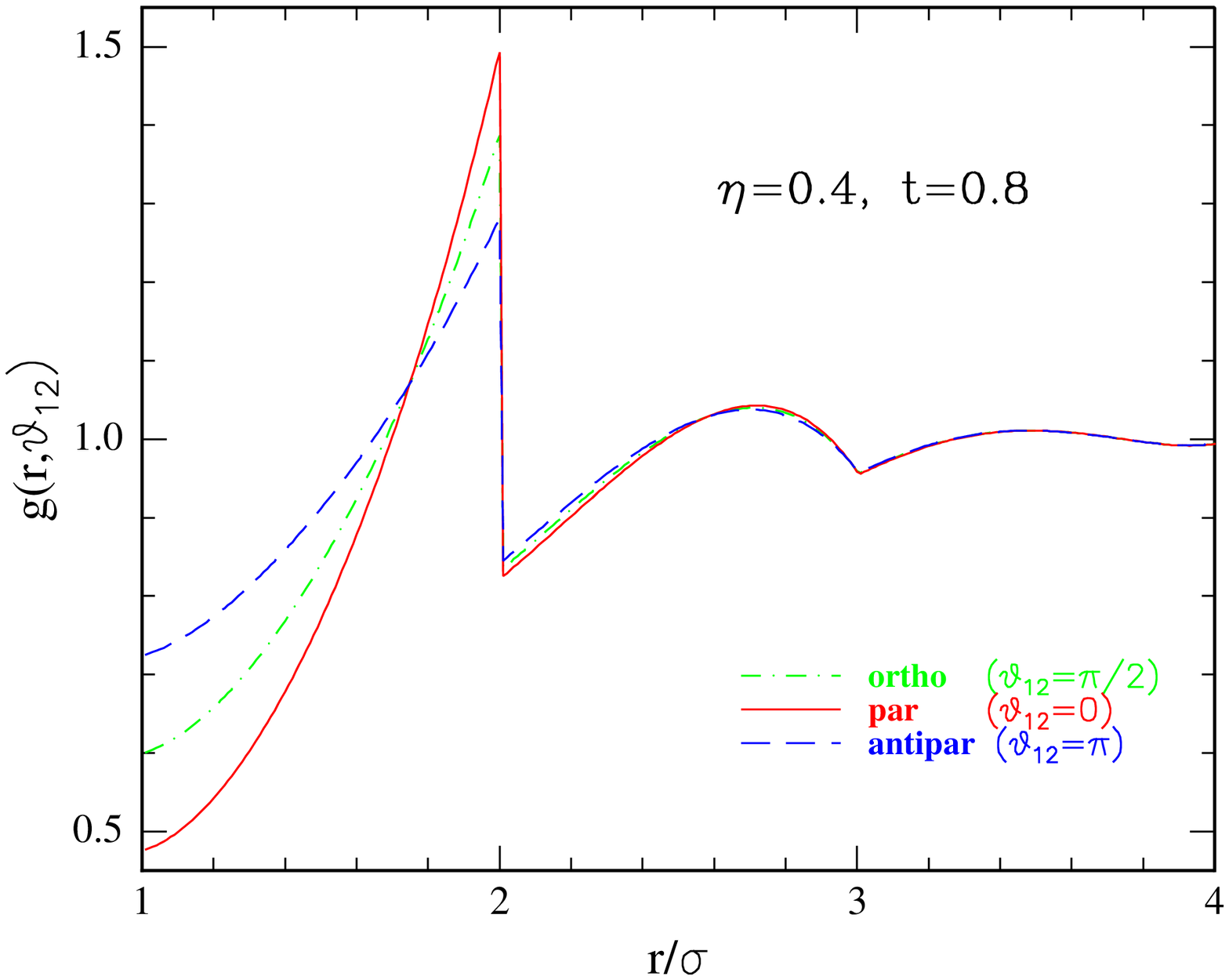}
\end{center}
\caption{(Color online) (a) Isotropic part of the pair correlation function, 
$g_{\mathrm{0}}(r)=g^{\mathrm{isoSHS}}(r)$, at $\protect\eta =0.4$, for $%
t=0,0.1,0.3,0.5,0.8$ corresponding to increasing adhesion strength or
decreasing temperature. $t=0$ yields the HS limit. (b) Behavior of $g(r,%
\protect\theta _{12}),$ when $\protect\alpha =1/2$, at $\protect\eta =0.4$
and $t=0.8,$ for three representative orientations $\protect\theta _{12}=0,%
\protect\pi /2,\protect\pi $ (parallel, orthogonal and anti-parallel
configurations). }
\label{fig1}
\end{figure}

Figure \ref{fig1}(a) depicts the behavior of $g_{0}(r)$, which coincides
with the reference isotropic part $g^{\mathrm{isoSHS}}(r)$ of the pair
correlation function, at $\eta =0.4$.

Here, $t=0$ gives the HS limiting case, $g^{\mathrm{HS}}(r)$, and we
consider increasing values of $t$, which correspond to increasing adhesion
or decreasing temperature, i.e. $t=0.1,0.3,0.5$ and $0.8$. The last $t$%
-value yields $\tau =1/(12t)\simeq 0.1$, which lies close to the critical
temperature of the isotropic fluid \cite{Miller04}.

Two features are noteworthy. First of all, the short-range interactions mainly
modify the short-range portions of the pair correlation functions. Very
pronounced effects are visible in the range $\sigma <r<2\sigma $, but
significant changes are also present all the way out to $r=4\sigma $ and
beyond, while the phase of the oscillations is clearly shifted by the
addition of the short-range attraction.

A second interesting feature concerns the $t$ dependence of $g_{0}(r)$ in
the first shell. As the adhesion strength increases from $t=0$ (HS) to $%
t=0.8 $, the contact value monotonically decreases, whereas a discontinuous
peak progressively builds up at $r=(2\sigma )^{-}$. This somewhat
counter-intuitive result can be easily understood in terms of the reduction
of the pressure exerted on particles 1 and 2 by the surrounding ones in the
presence of increasing attraction, thus providing an average larger
separation among 1 and 2.

Suppose now that we modulate this attraction with the anisotropic
dipolar-like dependence described above. When $\alpha=1/2$ the effect on $g$
is shown in Fig. \ref{fig1}(b) for three representative values of $%
\theta _{12}$: $\theta _{12}=0$ (parallel orientation), $\theta _{12}=\pi /2$
(orthogonal orientation) and $\theta _{12}=\pi $ (antiparallel orientation).
Note that in the orthogonal case the dipolar dependence vanishes and one
recovers the isotropic behavior. The main differences occur in the first
shell, where the orthogonal curve $\theta _{12}=\pi /2$ is bracketed between
the antiparallel ($\theta _{12}=\pi $) and the parallel ($\theta _{12}=0$)
results.

Similar qualitative results (with different separations among
parallel and antiparallel curves) are found when the angle between $\widehat{\mathbf{r}%
}_{12}$ and $\mathbf{u}_{1}$ is varied.

\begin{figure}[tbph]
\begin{center}
\includegraphics[width=12cm]{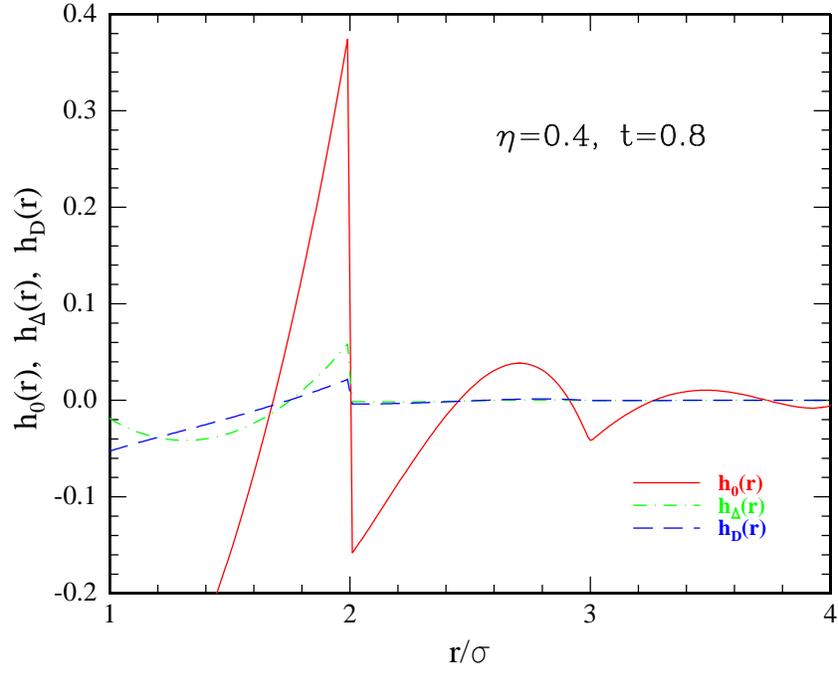} 
\end{center}
\caption{(Color online) The three components $h_{0}(r)$, $h_{\Delta }(r)$
and $h_{D}(r)$ of the molecular total correlation function, for $\protect%
\alpha =1/2$, $\protect\eta =0.4$, and $t=0.8$. At $r<\protect\sigma $ one
has $h_{0}(r)=-1$, $h_{\Delta }(r)=h_{D}(r)=0$. }
\label{fig2}
\end{figure}

From  Fig. \ref{fig1}(b) we note that at contact ($%
r=\sigma ^{+}$) the antiparallel configuration is more probable
that the nose-to-tail parallel one; conversely, at separations close to 
$r=2\sigma ^{-}$ the parallel alignment is predominant. 
This can also be confirmed by plotting the projections 
$h_{\Delta }(r)$ and $h_{D}(r)$ of the molecular correlation function
$h^{\mathrm{PY-OL}}(1,2)$ on the angular basis $\Delta(12)$ and $D(12)$ respectively.
This is depicted in Fig. \ref{fig2} where the isotropic corresponding contribution
$h_{0}(r)$ is also reported by contrast. One observes a weak negative correlation
for both quantities in the region $r \approx \sigma^{+}$ and, conversely,
a positive correlation close to $2 \sigma^{-}$. A crossing occurs approximately around the same
value $r \approx 1.7 \sigma$ where the parallel component in Fig. \ref{fig1}(b)
overtakes the antiparallel one, as expected.

As we shall see, however, this
is a \textit{local} ordering which does not affect the condensation process.

\begin{figure}[tbph]
\begin{center}
\includegraphics[width=12cm]{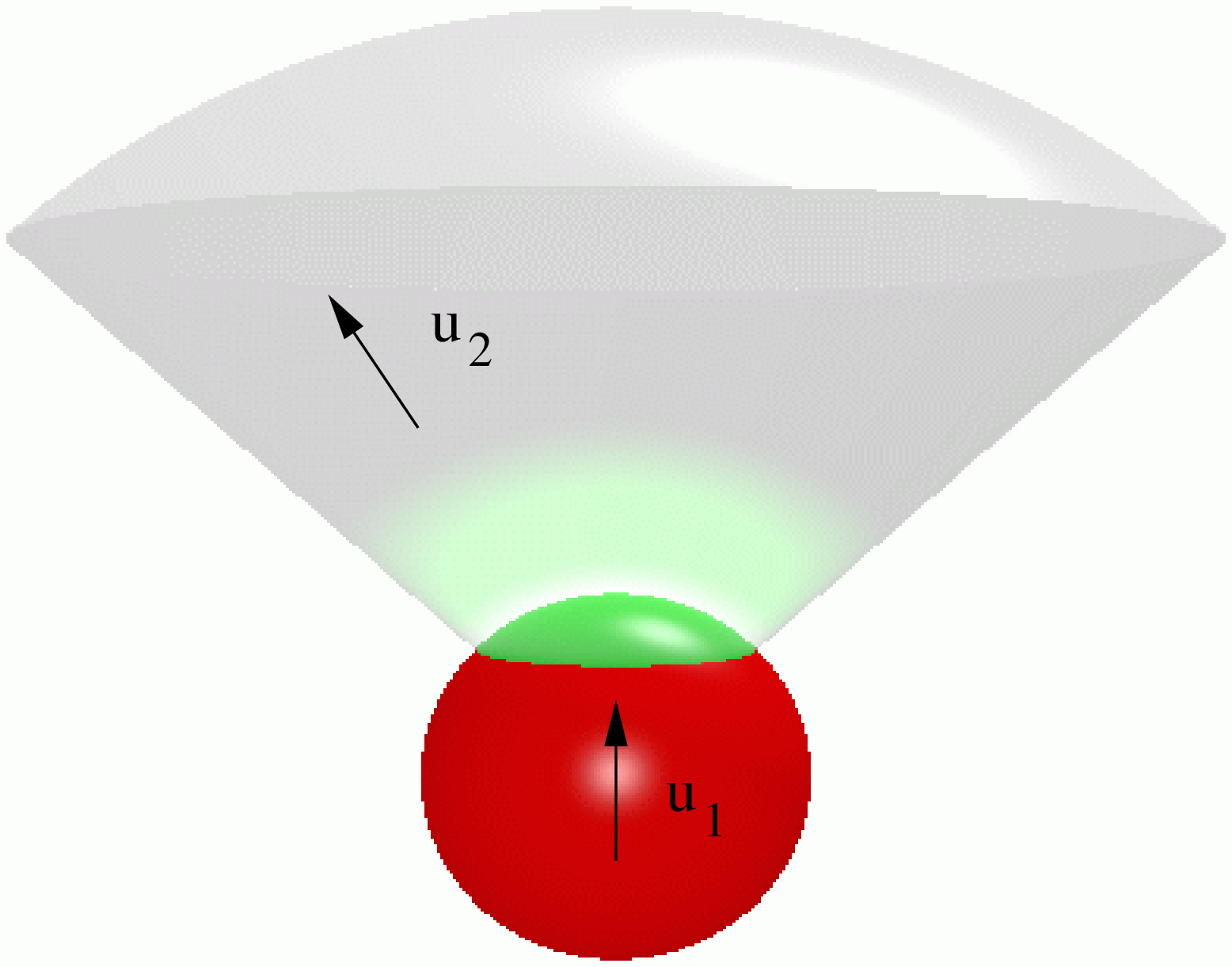} %
\includegraphics[width=12cm]{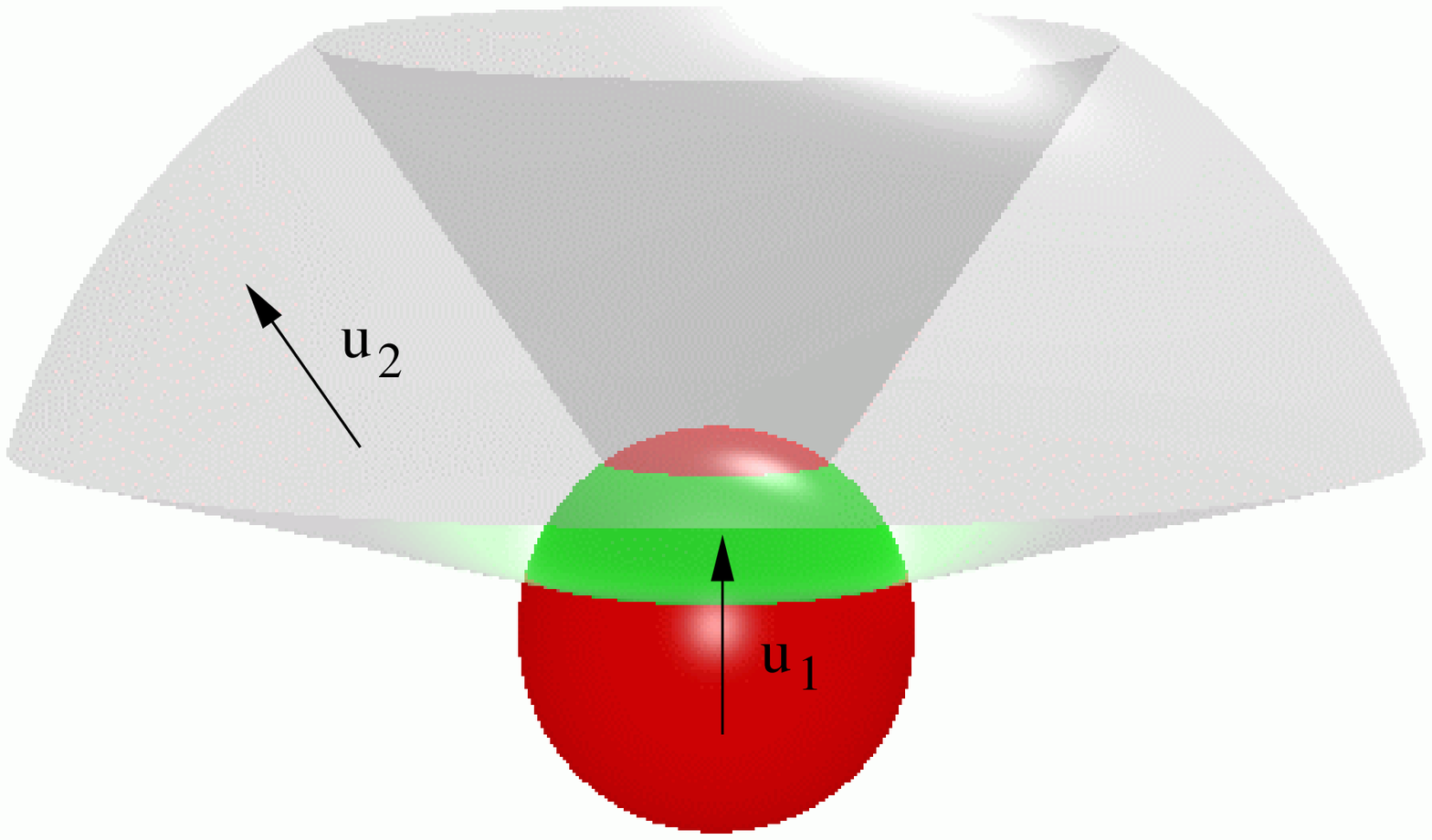}
\end{center}
\caption{(Color online) Schematic representation of the volume $V_{AB}$
(included in the shading area) in two different situations: (top panel) $\protect\theta %
_{A}=0$ giving the contribution from the forward region, and (bottom panel ) 
$\protect\theta _{A}\neq \protect\theta _{B}\neq 0$ giving information on
the lateral adjacent region.}
\label{fig3}
\end{figure}

In order to get more insight into such an orientational ordering, we compute
the number of particles with orientation $\mathbf{u}_{2}$ that a generic
reference particle 1 with orientation $\mathbf{u}_{1}$ `sees' in an
appropriate surrounding volume $V_{AB}$. Assuming $\mathbf{u}_{1}$ as polar
axis and taking into account the sphere $S$ with center $\mathbf{u}_{1}$ and
radius $R$, $V_{AB}$ is defined as the portion of $S$ corresponding to the
solid angle $\Omega _{AB}=\left\{ \left( \theta ,\varphi \right) \left\vert
\theta _{A}\leq \theta \leq \theta _{B},0\leq \varphi \leq 2\pi \right.
\right\} $ (see Fig. \ref{fig3}). Taking for instance $\theta _{A}=0$ and $%
\theta _{B}=\pi /3$, we can analyze the `forward ordering' as seen by the
reference particle, while choice $\theta _{A}=\pi /3$ and $\theta _{B}=\pi
/2 $ allows to discuss the `lateral ordering'.

The number of particles in an infinitesimal spherical cone of height $%
R=\lambda \sigma $ and infinitesimal solid angle $d\Omega _{r\text{ }}$ in a
given direction $\widehat{\mathbf{r}}$ is $d\mathcal{N}\left( \mathbf{u}_{1},%
\mathbf{u}_{2},\widehat{\mathbf{r}}\right) =d\Omega _{r}\int_{0}^{R}dr%
\mathbf{\ }r^{2}\ \rho g(1,2),$ where $d\Omega _{r}=d\widehat{\mathbf{r}}$.
In a finite solid angle $\Omega _{AB}$ 
\begin{equation}
\mathcal{N}\left( \mathbf{u}_{1},\mathbf{u}_{2}\right) =\int_{\Omega _{AB}}d%
\widehat{\mathbf{r}}_{\text{ }}\int_{0}^{R}dr\mathbf{\ }r^{2}\ \rho g(1,2)
\end{equation}

Using the first line of equation (\ref{model:eq5}) and equation (\ref{gPYOL}%
) we see that, within the PY-OL closure, $\left\langle g(12)\right\rangle _{%
\mathbf{u}_{1},\mathbf{u}_{2}}=g_{0}(r)$, so that the average number is 
\begin{equation}
\overline{\mathcal{N}}=\left\langle \mathcal{N}\left( \mathbf{u}_{1},\mathbf{%
u}_{2}\right) \right\rangle _{\mathbf{u}_{1},\mathbf{u}_{2}}=\rho \ \Omega
_{AB}\int_{0}^{R}dr\mathbf{\ }r^{2}\ g_{0}(r)=\rho \ \Omega _{AB}\ \sigma
^{3}\ I_{0}
\end{equation}%
with $\Omega _{AB}=\int_{0}^{2\pi }d\varphi \int_{\theta _{A}}^{\theta
_{B}}d\theta \ \sin \theta =2\pi (\cos \theta _{A}-\cos \theta _{B})$ and%
\begin{equation}
I_{0}=\left( \lambda ^{3}-1\right) /3+\int_{1}^{\lambda }dx\mathbf{\ }%
x^{2}h_{0\mathrm{,reg}}(x)+\Lambda _{0}
\end{equation}

Here we have used the results of Ref. I (see especially Section III D and
E), where $h_{0}(r)$ is decomposed into a `regular' term $h_{0\mathrm{,reg}%
}(r)$ and a `singular' term proportional to the delta function. A similar
decomposition is carried out (see again in Ref. I) for the $h_{\Delta }(r)$
and $h_{D}(r)$ parts. Using

\begin{equation}
\left( \Omega _{AB}\right) ^{-1}\int_{\Omega _{AB}}d\widehat{\mathbf{r}}\
D(1,2)=\ M_{AB}\left( \mathbf{u}_{1}\cdot \mathbf{u}_{2}\right) 
\end{equation}%
\begin{equation}
M_{AB}=\cos ^{2}\theta _{A}+\cos \theta _{A}\cos \theta _{B}+\cos ^{2}\theta
_{B}-1~,
\end{equation}%
we find that the fraction $X$ of particles with orientation $\mathbf{u}_{2}$
in the volume $V_{AB}$ around a reference particle having orientation $%
\mathbf{u}_{1}$, only depends upon the angle $\theta _{12}=\cos ^{-1}(%
\mathbf{u}_{1}\cdot \mathbf{u}_{2})$ and is given by 
\begin{equation}
X(\theta _{12})=\frac{\mathcal{N}\left( \mathbf{u}_{1},\mathbf{u}_{2}\right) 
}{\overline{\mathcal{N}}}=1+\frac{I_{\Delta }+M_{AB}\ I_{D}}{I_{0}}\ \left( 
\mathbf{u}_{1}\cdot \mathbf{u}_{2}\right)   \label{Xtheta}
\end{equation}%
\begin{equation}
\begin{array}{c}
I_{\Delta }=\int_{1}^{\lambda }dx\mathbf{\ }x^{2}h_{\Delta \mathrm{,reg}%
}(x)+\Lambda _{\Delta } \\ 
\\ 
I_{D}=\int_{1}^{\lambda }dx\mathbf{\ }x^{2}h_{D\mathrm{,reg}}(x)+\Lambda _{D}%
\end{array}%
\end{equation}

\begin{figure}[tbph]
\begin{center}
\includegraphics[width=12cm]{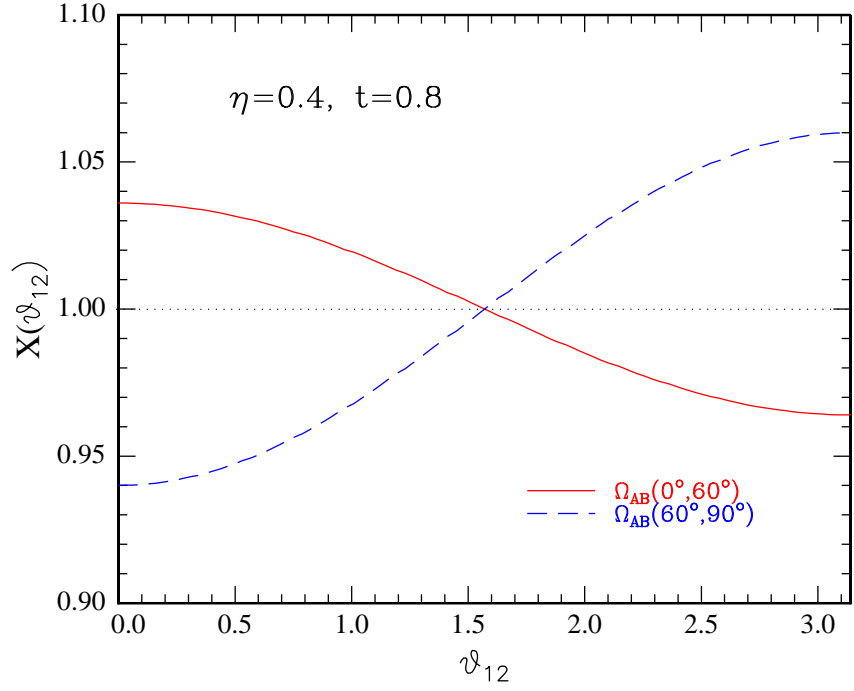} %
\includegraphics[width=12cm]{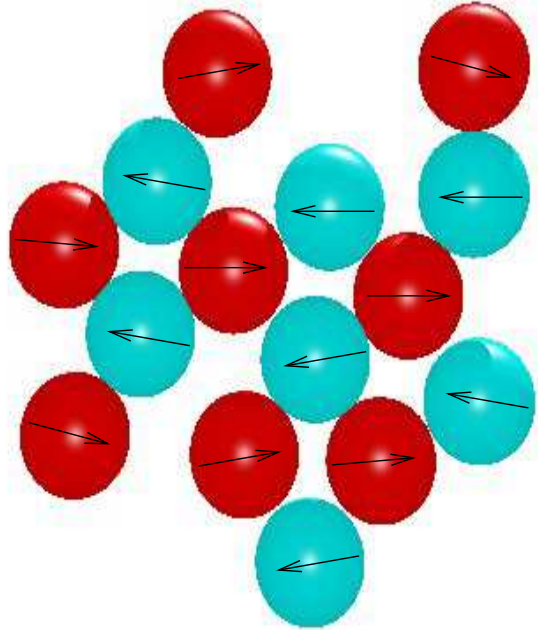}
\end{center}
\caption{(Color online) (a) Plot of the fraction $X$ of molecules with
orientation $\protect\theta _{12}$ contained in the volume $V_{AB\text{ }}$%
defined by $\protect\sigma \leq r\leq 2\protect\sigma $ and by the solid
angle $\Omega _{AB},$ with $\protect\theta _{A}=0,\protect\theta _{B}=%
\protect\pi /3$ for the forward direction and $\protect\theta _{A}=\protect%
\pi /3,\protect\theta _{B}=\protect\pi /2$ for the lateral direction.
Parameters are: $\protect\alpha =1/2$, $\protect\eta =0.4$ and $t=0.8$ in
all cases. (b) Schematic representation of a globular cluster, with
internal chain-like orientational ordering. }
\label{fig4}
\end{figure}

Fig. \ref{fig4} (a) depicts $X$ as a function of $\theta _{12}$ in the case $%
\lambda =2$ (first shell). In the forward region, represented by the solid
angle $\Omega _{AB}(0,\pi /3),$ we find $X(0)>X(\pi )$ so there are more
particles with parallel orientation, with respect to particle 1. On the
contrary, $X(0)<X(\pi )$ in the surrounding lateral region, characterized by 
$\Omega _{AB}(\pi /3,\pi /2)$, means that here the molecules with
antiparallel orientations prevail. Although these effects are rather small,
it is reasonable to expect that such differences should grow significantly
if the anisotropy parameter $\alpha t$ could become much larger than the
strength $t$ of isotropic adhesion.
Note that, while $X_{\mathrm{forward}}$ is larger than $X_{\mathrm{lateral}}$
in the interval $0\leq \theta _{12}\leq \pi /2$, an inversion occurs in the
region $\pi /2<\theta _{12}\leq \pi $ in agreement with the results of $%
g(r,\theta _{12})$ reported above (Fig \ref{fig1} (b)).

The above results are suggestive of the following physical picture.
Because of the limits imposed on the anisotropy parameter ($0\le \alpha \le 1/2$)
by the choice of the potential, the contribution of the dipolar-like
interaction is significantly weaker compared to the isotropic part,
and does not affect the main condensation process with the formation
of globule clusters of nearly isotropic shape.
This is in sharp contrast with the purely long-range dipolar models
which are mainly characterized by chain-like aggregation 
\cite{Tlusty00,Camp00,Ganzenmuller07}.
However a local ordering occurs within these globular agglomerates of condensation,
that are mainly formed by short portions of antiparallel chains running
next each other and held together essentially by the isotropic attraction.
This is schematically depicted in Fig. \ref{fig4}(b).

We note that only particles belonging to different,
antiparallel, chains have \textit{direct contact}. Consecutive
molecules with parallel noise-to-tail orientation -- i.e. belonging to the
same chain -- are not in contact, but lie with average separations slightly smaller
than $2\sigma $ as suggested by the behavior of $g$ in Fig.\ref{fig1}(b). 
Thus antiparallel molecules of adjacent chains 'mediate' an
\textit{indirect contact} between consecutive particles of a given chain.

Once again, we stress that this phenomena should be considered a local
fluctuation with very short range (of the order of one shell, as remarked)
and does not extend to the entire fluid. This can be readily checked by
considering the limit $\lambda \rightarrow \infty $, in which case one finds 
$I_{D}=0$, so that the dependence from $\mathbf{u}_{1}\cdot \mathbf{u}%
_{2}=\cos \theta _{12}$ is averaged to zero. As we shall see below, a direct
consequence of this is that the coexistence line of the isotropic model is
not significantly affected by the anisotropic part, within the PY-OL
approximation.

In view of the last remark, one might rightfully wonder whether the
anisotropic part plays any role in the thermodynamics of our model. We can
convince ourselves that the answer is positive, by considering the exact
third virial coefficient as defined by 
\begin{equation}
B_{3}=-\frac{1}{3V}\int d\mathbf{r}_{1}~d\mathbf{r}_{2}~d\mathbf{r}_{3}\
\left\langle f\left( 1,2\right) f\left( 1,3\right) f\left( 2,3\right)
\right\rangle _{\mathbf{u}_{1},\mathbf{u}_{2},\mathbf{u}_{3}}
\end{equation}

Note that, in view of Eq. (\ref{model:eq5}), the exact second virial
coefficient $B_{2}=-\frac{1}{2}\int d\mathbf{r\ }\left\langle
f(1,2)\right\rangle _{\mathbf{u}_{1},\mathbf{u}_{2}}$ coincides with its
isotropic counterpart. However, this is not the case for $B_{3}$, that can
be computed following the method outlined in Ref. \onlinecite{Fantoni07} for
patchy sticky hard spheres, a close relative to the present model. One finds 
\begin{equation}
b_{3}=B_{3}/v_{0}^{2}=10-60t\chi _{1}+144t^{2}\chi _{2}-96t^{3}\chi _{3}~,
\end{equation}%
where $v_{0}=\left( \pi /6\right) \sigma ^{3}$ and 
\begin{equation}
\begin{array}{c}
\chi _{1}=\langle \epsilon (1,2)\rangle _{\mathbf{u}_{1},\mathbf{u}_{2}}%
\text{ \ \ \ \ \ \ \ \ \ \ }\chi _{2}=\langle \epsilon (1,2)\epsilon
(1,3)\rangle _{\mathbf{u}_{1},\mathbf{u}_{2},\mathbf{u}_{3}} \\ 
\\ 
\chi _{3}=\langle \epsilon (1,2)\epsilon (1,3)\epsilon (2,3)\rangle _{%
\mathbf{u}_{1},\mathbf{u}_{2},\mathbf{u}_{3}}~.%
\end{array}
\label{3rd_2}
\end{equation}%
Again using (\ref{model:eq5}), we find $\chi _{1}=1=\chi _{2}$. The exact
value of $\chi _{3}$ turns out to be 
\begin{equation}
\chi _{3}=1-\frac{11}{72}\alpha ^{3}~.  \label{3rd_3}
\end{equation}%
The anisotropic contribution is represented by the term $-(11/72)\alpha
^{3}\approx 0.02$, that is very weak with respect to the isotropic one.

Having assessed the limits of the model, we now turn to discuss the limits
of the approximation involved in the PY-OL closure. A simple and direct way
to quantify its deviation from the exact results is to consider the first
order density expansion of the exact direct correlation function $%
c(1,2)=f(1,2)+c^{(1)}(1,2)\rho +\cdots $. We find 
\begin{equation}
c^{(1)}(1,2)=c_{\mathrm{PY-OL}}^{(1)}(1,2)+c_{\mathrm{ex}}^{(1)}(1,2)
\label{c1}
\end{equation}%
with $c_{\mathrm{PY-OL}}^{(1)}(1,2)=c_{0}^{(1)}(r)+c_{\Delta }^{(1)}(r)\
\Delta (1,2)+c_{D}^{(1)}(r)\ D(1,2)$, and 
\begin{equation}
c_{\mathrm{ex}}^{(1)}(1,2)=(\alpha t)\ \left[ \ \gamma _{\Delta
}^{(1)}(\sigma )\ \Delta (1,2)D(1,2)+\gamma _{D}^{(1)}(\sigma )\ D^{2}(1,2)%
\right] \ \sigma \delta (r-\sigma ).
\end{equation}%
Whereas the PY closure includes both $c_{\mathrm{PY-OL}}^{(1)}(1,2)$ and $c_{%
\mathrm{ex}}^{(1)}(1,2)$, thus reproducing the exact third virial
coefficient through $B_{3}=-\frac{1}{3}\int d\mathbf{r\ }\langle
c^{(1)}(1,2)\rangle _{\mathbf{u}_{1},\mathbf{u}_{2}}$, the PY-OL
approximation omits the contributions included in $c_{\mathrm{ex}%
}^{(1)}(1,2) $. Consequently, $b_{3}^{\mathrm{PY-OL}}$ reduces to the purely
isotropic contribution: $b_{3}^{\mathrm{iso}}=10-60t+144t^{2}-96t^{3}$. The
anisotropic contribution $b_{3}^{\mathrm{aniso}}=\frac{44}{3}(\alpha t)^{3}$%
, stemming from $c_{\mathrm{ex}}^{(1)}(12)$, can be easily computed again
with the help of Eqs. (\ref{model:eq5}), in agreement with the exact result Eq.(\ref{3rd_3}).

Next we consider the thermodynamics. As in all approximate closures, even
within the PY one there exist three standard routes to the equation of
state: compressibility, energy, and virial routes.

In the first two cases, it is easy to convince oneself that the result is
the same as for the isotropic SHS system calculated in Ref. \onlinecite{Baxter68}.
This is again due to Eq. (\ref{model:eq5}) and is a consequence of the
linearity of the expansion in the angular part involved in the PY-OL
approximation, Eq. (\ref{summary:eq2}), and of the minor role played by the
anisotropic part, as testified by the weak $\alpha -$ dependence of the
third virial coefficient (Eqs. (\ref{3rd_2}) and (\ref{3rd_3})). This is
also in agreement with the stability analysis of Ref. I, which can also be
extended to finite values of the wave vector $k$.

As often the case, the virial route is more delicate. Here, standard
steps lead to 
\begin{equation}
\frac{\beta p}{\rho }=1+4\eta y_{0}(\sigma )-4\eta t\left\{ 2y_{0}(\sigma
)+Y_{0}^{\prime }(\sigma )+\frac{2\alpha }{3}[2y_{D}(\sigma )+Y_{D}^{\prime
}(\sigma )]\right\}  \label{pvirial}
\end{equation}%
where $\eta =\rho v_{0}$ is the packing fraction, and $%
Y_{0}(r)=ry_{0}(r),Y_{D}(r)=ry_{D}(r)$, with $y^{\mathrm{PY}}(1,2)=1+\gamma
(1,2)$ being the PY cavity function.

\begin{figure}[tbph]
\begin{center}
\includegraphics[width=12cm]{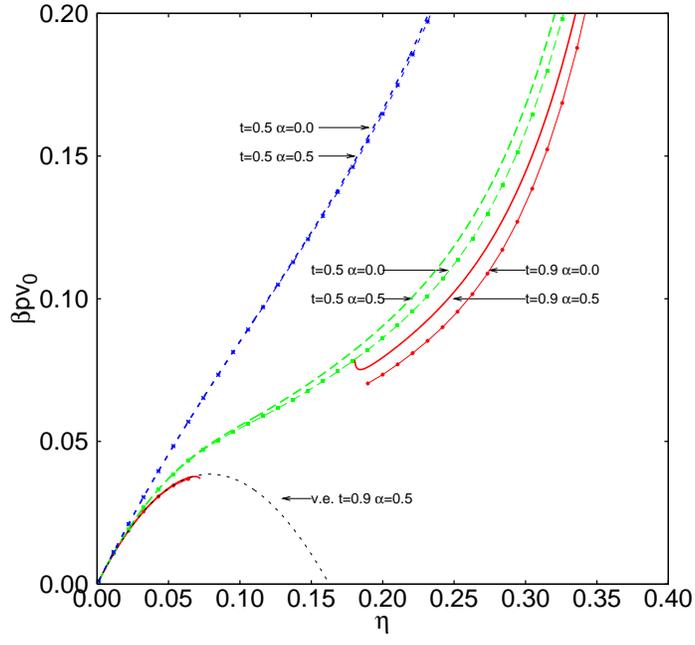}
\end{center}
\caption{(Color online) A plot of $\protect\beta pv_{0}$ versus the packing
fraction $\protect\eta $ for four different values of $t=1/(12\protect\tau %
), $ with ($\protect\alpha =1/2$) and without ($\protect\alpha =0$) the
anisotropic contribution, using the virial route to the pressure. The 3rd
order virial expansion is also added (v.e.) in the most relevan case $t=0.9$ and $\alpha=0.5$ for comparison. 
The part of the lines which are not shown correspond to a loss of solutions.}
\label{fig5}
\end{figure}

For given $t,\eta $, one can calculate $y_{0}(\sigma )$, $y_{D}(\sigma )$, $%
Y_{0}^{\prime }(\sigma )$ and $Y_{D}^{\prime }(\sigma )$ analytically using
expressions from Ref. I. However, $Y_{0}^{\prime }(\sigma )$ and $%
Y_{D}^{\prime }(\sigma )$ require some care, since space derivative and
sticky limit do not commute \cite{Baxter68}. So from Eq. (\ref{pvirial}) one
finds the virial pressure. The corresponding results are collected in Fig. %
\ref{fig5} for different values of $t$, at both $\alpha =0$ (isotropic case)
and $\alpha =1/2$ (with the anisotropic contribution included). A comparison
with the virial expansion up to the third virial coefficient is also added
in the case $t=0.9$ and $\alpha =1/2$.

In agreement with the previous structural findings, we find a dependence on
the anisotropy. This is very small for $t \le 0.5$ but increasingly appreciable
for larger values of the adhesion strength $t$. In Fig. \ref{fig5}
one clearly sees that the pression increases by roughly $10\%$ on going from
$\alpha =0$ (no anisotropy) to $\alpha =0.5$ (maximum anisotropy) for
$t=0.9$ and $\eta \ge 0.2$.

\begin{figure}[tbph]
\begin{center}
\includegraphics[width=12cm]{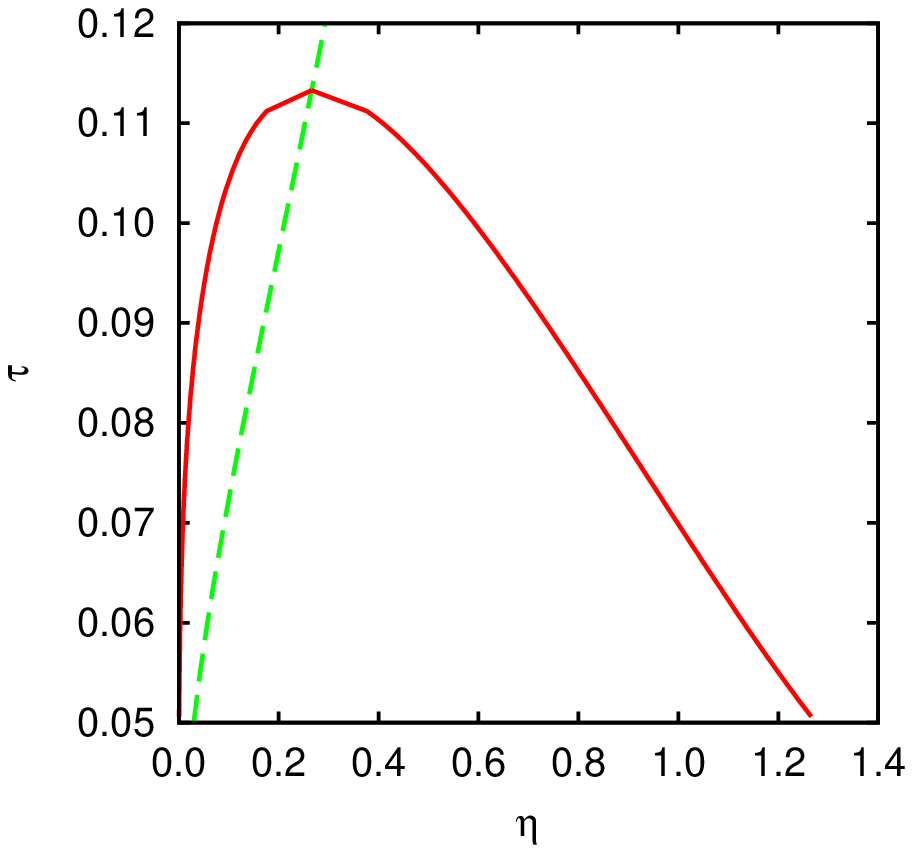}
\end{center}
\caption{(Color online) Phase diagram of our dipolar-like sticky hard
spheres, calculated for defect energies $\protect\epsilon _{1}=0.186$ and $%
\protect\epsilon _{3}=-0.0102$. At the critical point the binodal curve
(continuous line) and the connectivity transition (dashed curve) meet. The
line denote the end-rich \textquotedblleft gas\textquotedblright\ with the
junction-rich \textquotedblleft liquid\textquotedblright .}
\label{fig6}
\end{figure}

Despite the strong differences with the pure dipolar case, it proves
instructive to get some insight into the competition between the tendency to
condensation on the one hand and to chaining on the other hand, by applying
to the present model the arguments put forward by Tlusty and Safran \cite%
{Tlusty00} in the dipolar case. These authors devised a phenomenological
theory, where the two above-mentioned tendencies are represented by the
concentrations of \ `junctions' and `ends', respectively (see Ref. \onlinecite%
{Tlusty00}). We have closely followed their arguments to derive the critical parameters 
$\tau _{c}$ and $\eta _{c}$ of
the present model in terms of the energies $\epsilon _{1}$, $\epsilon _{3}$
of ends and junctions respectively. One finds \cite{Tlusty00} 
\begin{equation}
\tau _{c}=\frac{\epsilon _{1}-3\epsilon _{3}}{3\ln 3-2\ln 2}~,\text{ \ \ \ \
\ \ \ \ \ \ \ }\ln \eta _{c}=-\frac{\epsilon _{1}(2\ln 3-\ln 2)-\epsilon
_{3}\ln 2}{\epsilon _{1}-3\epsilon _{3}}~.
\end{equation}%
which coincides with the results of Ref. \onlinecite{Tlusty00}. Matching this
critical values with the one of the isotropic adhesive spheres of Miller and
Frenkel \cite{Miller04}, $\tau _{c}=0.1133$ and $\eta _{c}=0.266$, we find $%
\epsilon _{1}=0.186$ and $\epsilon _{3}=-0.0102$ (the value $\epsilon _{3}<0$
means that junctions are enhanced with respect to ends, once more favoring
condensation). In our results the number of ends $\rho _{1}$ and junctions $%
\rho _{3}$ turn out to be equal ($\rho _{1}=\rho _{3}$) at the critical
point, which is thus a point of connectivity transition in the system. Fig. %
\ref{fig6} depicts the coexistence line, which does not display the
re-entrance characteristic of the pure dipolar case (compare with  Fig. 2 of Ref. 
\onlinecite{Tlusty00}). This is in complete agreement with the remark by Tlusty and
Safran that the addition of an isotropic short-range attraction \ -- such as
the case of the present model -- reports the curve to its characteristic
parabolic shape (see also Fig. 3 of Ref. \onlinecite{Tlusty00}).
This is also consistent with very recent numerical simulation results \cite%
{Ganzenmuller09,Martin09} showing that the addition of a very weak
isotropic attraction to the dipolar HS potential makes the condensation
transition easily observable.

In summary, we have studied structural and thermophysical properties of a
particular hard-core fluid where the attractive part of the potential
includes an anisotropy of dipolar form infinitesimally short and infinitely
strong.

Any two molecules of the fluid interact only at contact with a potential
having, in addition to an adhesive isotropic part of the Baxter type, an
additional adhesive term, whose intensity depends upon the mutual
orientation of the two particles in a dipolar fashion.

Our potential belongs to a class of simple anisotropic models that
have recently attracted considerable interest in connection with
aggregation phenomena in colloidal fluids, polymers and globular proteins,
because of their possible experimental relevance for self-assembling
materials and biological viruses.

The extremely short-range nature of this peculiar dipolar interaction
strongly contrasts with the long-range nature of the dipolar hard sphere
model. In the latter case, the formation of chains and long anisotropic
agglomerates significantly affects the possibility of a gas-liquid
transition. Using a simplified treatment of the angular part, based upon a
first-order expansion in angular invariants so to allow an almost fully
analytical solution, we have shown that only the local (first few)
coordination shells are affected by the anisotropy. This is due to
the fact that the the orientationally dependent part of the
potential has a relatively weak strength with respect to the isotropic
attractive term, as forced by the particular choice of the potential
associated with the $0 \le \alpha \le 1/2$ limits.
As a result, all structural and thermodynamical properties are only mildly affected by the
anisotropic adhesion.

Nonetheless, the competition of the two adhesive terms (the isotropic and the
anisotropic ones) gives rise to an anisotropic local ordering within
each (almost isotropic) molecular agglomerate 
consisting of short chains of molecules
with parallel head-to-tail orientation, `glued' to similar chains globally
oriented in the opposite direction, thus giving an antiparallel alignment
for particles belonging to two adjacent chains.

It would be interesting to contrast the present results
with more realistic models incorporating a competition
between an isotropic and anisotropic short range interactions,
such as for instance Stockmayer fluids \cite{Leeuwen93}, 
dipolar Yukawa HS fluids \cite{Slazai99} or combination
of dipolar and square-well potentials \cite{Martin09}.

In spite of its simplicity, the results of the present work suggest that, in the
presence of dipolar-like anisotropy, one can continuously tune from
situations only affecting the local ordering (such as in the case presented
here) to situations where this effect is much more global (such as the real
dipolar case), by simply adjusting the range of interaction.

\begin{acknowledgments}
 Funding from PRIN-COFIN 2007 is gratefully acknowledged.
\end{acknowledgments}

%%%%%%%%%%%%%%%%%%%%%% Bibliography %%%%%%%%%%%%%%%%%%%%%%%%%%%%%%%%%%%%
%\references

\bigskip

%%%%%%%%%%%%%%%%%%%%%%% Figures %%%%%%%%%%%%%%%%%%%%%%%%%%%%%%
\newpage

% FIG 1
%
%

\clearpage 
% FIG 2
%
%

\clearpage
% FIG 3
%

\clearpage 
% FIG 4
%

\clearpage 
% FIG 5
%

\clearpage 
% FIG 6
%

%%%%%%%%%%%%%%%%%%%%%%%%%%%%%%%%%%%%%%%%%%%%%%%%%%%%%%%%%%%%%%%%%%%%%%%%%%%%%%%%%%%%%%%%%%%%%%%%%%%%%%%%%%%%%%%%%%%%%%%%%%%%%%%%%%%%%%%%%%%%%%%%%%%%%%%%%%%%%%%%%%%%%%%%%%%%%%%%%%%%%%%%%%%%%%%%%%%%%%%%%%%%%%%%%%%%%%%%%%%%%%%%%%%%%%%%%%%%%%%%


\begin{thebibliography}{99}

\bibitem{Zaccarelli09} E. Zaccarelli, C. Valeriani, E. Sanz, W.C.K. Poon,
M.E. Cates, and P. N. Pusey, Phys. Rev. Lett. \textbf{103}, 135704 (2009)

\bibitem{Zaccarelli08} E. Zaccarelli, J. Phys.: Condensed Matter \textbf{19},
323101 (2007).

\bibitem{Yethiraj03} A. Yethiraj and A. van Blaaderen, Nature \textbf{421},
513 (2003).

\bibitem{Lu08} P.J. Lu, E. Zaccarelli, F. Ciulla, A. B. Schofield, F.
Sciortino and D. Weitz, Nature \textbf{453}, 499 (2008).

\bibitem{Kern03} N. Kern, and D. Frenkel, J. Chem. Phys. \textbf{118}, 9882
(2003).

\bibitem{Fantoni07} R. Fantoni, D. Gazzillo, A. Giacometti, M.A. Miller and
G. Pastore, J. Chem. Phys. \textbf{127}, 234507 (2007); A. Giacometti, F. Lado,
J. Largo, G. Pastore and F. Sciortino, J. Chem. Phys. \textbf{131}, 174114 (2009)

\bibitem{Liu07} H. Liu, S. K. Kumar and F. Sciortino, J. Chem. Phys. \textbf{%
127}, 084902 (2007).

\bibitem{Foffi07} G. Foffi and F. Sciortino, J. Phys. Chem. B \textbf{111},
9702 (2007).

\bibitem{Bianchi08} E. Bianchi, P. Tartaglia, E. Zaccarelli and F.
Sciortino, J. Chem. Phys. \textbf{128}, 144504 (2008).

\bibitem{Gogelein08} C. G\"{o}gelein, G. N\"{a}gele, R. Tuinier, T. Gibaud,
A. Strader and P. Schurtenberger, J. Chem. Phys. \textbf{129}, 085102 (2008).

\bibitem{Tavares09} J.M. Tavares, P.I.C. Teixeira, and M.M. Telo de Gama,
Mol. Phys. \textbf{107}, 453 (2009).

\bibitem{Wertheim71} M. S. Wertheim, J. Chem. Phys. \textbf{55}, 4291 (1971).

\bibitem{Tlusty00} T. Tlusty and S.A. Safran, Science \textbf{290}, 1328
(2000).

\bibitem{Camp00} P.J. Camp, J.C. Shelley and G.N. Patey, Phys. Rev. Lett. 
\textbf{84}, 115 (2000).

\bibitem{Ganzenmuller07} G. Ganzenm\"{u}ller and P.J. Camp, J. Chem. Phys. 
\textbf{126}, 191104 (2007).

\bibitem{Baxter68} R. J. Baxter, J. Chem. Phys. \textbf{49}, 2770 (1968).

\bibitem{Gazzillo08} D. Gazzillo, R. Fantoni, and A. Giacometti, Phys. Rev.
E \textbf{78}, 021201, (2008).

\bibitem{Jackson88} G.\ Jackson, W. G. Chapman, and K. E. Gubbins, Mol.
Phys. \textbf{65}, 1 (1988).

\bibitem{Sear99} R. P. Sear, J. Chem. Phys. \textbf{111}, 4800 (1999).

\bibitem{Ghonasgi95} D. Ghonasgi, and W. G. Chapman, J. Chem. Phys. \textbf{%
102}, 2585 (1995).

\bibitem{Mileva00} E. Mileva, and G. T. Evans, J. Chem. Phys. \textbf{113},
3766 (2000).

\bibitem{Zhang05} Z. Zhang, A. S. Keys, T. Chen, and S. C. Glotzer,
Langmuir, \textbf{21}, 11547 (2005).

\bibitem{Gray84} C. G. Gray, and K. E. Gubbins, \textit{Theory of Molecular
Fluids, Vol. I}, Appendix 3E (Clarendon Press, Oxford, 1984).

\bibitem{Miller04} M. Miller and D. Frenkel, J. Chem. Phys. \textbf{121},
535 (2004).

\bibitem{Ganzenmuller09} G. Ganzenm\"{u}ller, G.N. Patey, and P.J. Camp,
Mol. Phys. \textbf{107}, 403 (2009).

\bibitem{Martin09} M. Mart\'{\i}n-Betancourt, J.M. Romero-Enrique, and L.F.
Rull, Mol. Phys. \textbf{107}, 563 (2009).

\bibitem{Leeuwen93} M.E. van Leeuwen, and B. Smit, Phys. Rev. Lett. \textbf{%
71}, 3991 (1993).

\bibitem{Slazai99} I. Slazai, D. Henderson, D. Boda, and K.Y. Chan, J. Chem.
Phys. \textbf{119}, 337 (1999).
\end{thebibliography}
\end{document}